\documentclass[prd,aps,nofootinbib,twocolumn,superscriptaddress]{revtex4}
\usepackage{epsfig}
\usepackage{amsmath, slashed}
\usepackage{xcolor}
\usepackage{appendix}
\usepackage{dsfont}

\usepackage[normalem]{ulem}

\usepackage{bm}
\usepackage{amsfonts}

\definecolor{pine}{rgb}{0.3, 0.5, 0.3}

\allowdisplaybreaks
\begin{document}
%\preprint{xxx}\preprint{xxx}

\title{Spin-orbit correlation and  spatial distributions for spin-0 hadrons}

\author{C\'edric Lorc\'e}
\email[Corresponding author: ]{cedric.lorce@polytechnique.edu}
\affiliation{CPHT, CNRS, \'Ecole polytechnique, Institut Polytechnique de Paris, 91120 Palaiseau, France}

\author{Qin-Tao Song}
\email[Corresponding author: ]{songqintao@zzu.edu.cn}
\affiliation{School of Physics, Zhengzhou University, Zhengzhou, Henan 450001, China}
%\pacs{12.15.-y; 12.38.-t; 12.39.St; 13.85.-t; 13.88.+e}

\date{\today}

\begin{abstract}
{The spin-orbit correlation in spin-0 hadrons can be investigated through the kinetic energy-momentum tensor form factor $\tilde F^q(t)$. We observe that the latter is also related to a torque about the radial direction, which we interpret as a chiral stress. If we neglect the quark mass contribution, then $\tilde F^q(t)$ is simply proportional to the electromagnetic form factor for spin-0 hadrons, and the spin-orbit correlation is equal to minus half of the valence quark number. Given the extensive studies on the electromagnetic form factor for spin-0 hadrons such as pions, kaons, and the $\alpha$ particle, we present the spatial distributions of chiral stress and kinetic
spin-orbit correlation based on current parametrizations of the pion electromagnetic form factor.}

\end{abstract}

\maketitle

\section{Introduction}
\label{sect1}
Spin-0 hadrons play a crucial role in particle physics, nuclear physics and cosmology. For example, the pion meson is one of the Goldstone bosons, and it is also considered as the carrier particle of the nuclear force that binds nucleons together within the nucleus. In addition, the scalar nucleus  such as  $^4$He ($\alpha$ particle) and the newly observed anti-$\alpha$ particle~\cite{STAR:2011eej} are used to study the matter and antimatter asymmetry in the universe. Thus, it is of top priority to study the inner structures of spin-0 hadrons. In that respect, electromagnetic (EM) form factors (FFs) play a key role and have attracted a lot of attention. In particular, there have been numerous experimental and theoretical studies on the EM FFs for pions~\cite{JeffersonLabFpi:2007vir,JeffersonLabFpi-2:2006ysh,JeffersonLab:2008gyl,JeffersonLab:2008jve,Perry:2018kok,Gao:2021xsm, Cui:2021aee,Lepage:1979zb,Chang:2013nia,Miller:2009qu,Alberg:2024svo,Brandt:2013dua,ETM:2017wqc,Fukaya:2014jka,Aoki:2015pba,Koponen:2015tkr,Feng:2019geu,Wang:2020nbf,Chen:2023byr,Bruch:2004py,CLEO:2005tiu,Seth:2012nn,Xu:2023izo}, kaons~\cite{Bruch:2004py,CLEO:2005tiu,Seth:2012nn,Xu:2023izo,CMD-2:2008fsu,BaBar:2015lgl,CMD-3:2016nhy,BESIII:2023zsk,Wu:2008yr,Gao:2017mmp,Krutov:2016luz,Stamen:2022uqh,Ahmed:2023zkk,Miramontes:2022uyi,Chen:2024oem}, and the $\alpha$ particle~\cite{Schiavilla:1990zz,JeffersonLabHallA:2013cus,Arnold:1978qs,Doyle:1992zz,Carlson:1997qn,Bacca:2014tla}.

In addition to the EM FFs, one can also
investigate the energy-momentum tensor (EMT) FFs~\cite{Kobzarev:1962wt,Pagels:1966zza} using generalized parton distributions (GPDs) of hadrons. The EMT FFs help us reveal the spin, mass, and mechanical structures of hadrons~\cite{Ji:1996ek,Burkert:2023wzr,Burkert:2018bqq,Freese:2021czn,Lorce:2014mxa,Leader:2013jra,Polyakov:2002yz,Polyakov:2018zvc,Lorce:2017wkb,Lorce:2018egm,Bhoonah:2017olu,Ji:2020ena,Kumericki:2019ddg,Cao:2023ohj,Shanahan:2018nnv,Owa:2021hnj,Kim:2022wkc, Chakrabarti:2020kdc}. The hadron GPDs can be accessed via some exclusive reactions such as deeply virtual Compton scattering (DVCS).
For the $\alpha$ particle, the DVCS reaction $\gamma^{*}$\,$^4\text{He} \to \gamma$\,$^4\text{He} $ has recently been measured by JLab in 2017~\cite{CLAS:2017udk} and 2021~\cite{CLAS:2021ovm}.
Since at the moment there are no facilities capable of directly measuring the DVCS reactions for scalar mesons, it has been proposed to study instead the $\gamma^*\gamma \to M \bar{M}$ reaction to extract the meson generalized distribution amplitudes (GDAs), from which one can deduce the EMT FFs in the timelike region~\cite{Diehl:2003ny, Muller:1994ses, Diehl:1998dk,Polyakov:1998ze}.
 Recently, Belle has measured  the cross sections for the production of a pion pair~\cite{Belle:2015oin} and a kaon pair~\cite{Belle:2017xsz}. Subsequently, the pion GDA and EMT FFs were extracted through the analysis of the experimental cross section~\cite{Kumano:2017lhr}. In the future, the pion GPDs could also be accessed indirectly at the Electron-Ion Collider, via a process where the pion is emitted by a proton~\cite{Amrath:2008vx, Chavez:2021koz, Chavez:2021llq}.

In this work, we study the matrix elements of the \textsf{P}-odd EMT for spin-0 hadrons. The \textsf{P}-odd
EMT provides the physical information about the left- and right-handed quarks separately, and is in particular directly related to the notion of quark spin-orbit correlation $\langle S^q_z L^q_z \rangle$ inside the hadrons~\cite{Lorce:2011kd,Lorce:2014mxa,Kim:2024cbq,Hatta:2024otc,Bhattacharya:2024sno,Tan:2021osk, Acharyya:2024enp}.

\section{Spin-orbit correlation in spin-0 hadrons}
\label{sect2}

The kinetic EMT for the quark flavor $q$ is given by
\begin{equation}\label{localEMT}
   \hat{T}^{\mu\nu}_q(x)= \overline\psi(x)\gamma^\mu \tfrac{i}{2}\overset{\leftrightarrow}{D}\!\!\!\!\!\phantom{D}^\nu\psi(x),
\end{equation}
where  $\overset{\leftrightarrow}{D}\!\!\!\!\!\phantom{D}^\nu=\overset{\rightarrow}{\partial}\!\!\!\!\phantom{\partial}^\nu-\overset{\leftarrow}{\partial}\!\!\!\!\phantom{\partial}^\nu-2ig A^\nu(x)$ is the  covariant derivative. Taking the difference between the  right- and left-handed quark contributions, one arrives at the following expression
\begin{align}
 \hat{T}^{\mu\nu}_{q5}(x)&=
 \overline\psi_R(x)\gamma^\mu \tfrac{i}{2}\overset{\leftrightarrow}{D}\!\!\!\!\!\phantom{D}^\nu\psi_R(x)-
  \overline\psi_L(x)\gamma^\mu \tfrac{i}{2}\overset{\leftrightarrow}{D}\!\!\!\!\!\phantom{D}^\nu\psi_L(x) \nonumber \\
  &=   \overline\psi(x)\gamma^\mu \gamma_5 \tfrac{i}{2}\overset{\leftrightarrow}{D}\!\!\!\!\!\phantom{D}^\nu\psi(x),\label{localEMTg5}
   \end{align}
known as the \textsf{P}-odd EMT.

Similarly to the quark kinetic orbital angular momentum (OAM) operator, the quark kinetic spin-orbit correlation operator is defined on the light front as~\cite{Lorce:2014mxa}
  \begin{equation}\label{opsl}
\hat{C}^q_z= \int dx^-d^2x_{\scriptscriptstyle{T}}\left[x^1\hat{T}^{+2}_{q5}(x) - x^2\hat{T}^{+1}_{q5}(x)\right],
\end{equation}
and can be interpreted as the difference of longitudinal OAM between right- and left-handed quarks. For convenience, we introduced two light-front vectors $n$ and $\bar{n}$,
\begin{equation}\label{lcvs}
n^{\mu}=\tfrac{1}{\sqrt{2}}(1,0,0,-1),\qquad  \bar{n}^{\mu}=\tfrac{1}{\sqrt{2}}(1,0,0,1).
\end{equation}
The light-front components of a vector $a$ are then given by $a^+=a \cdot n$ and $a^-=a \cdot \bar{n}$, and the vector $a$ can be re-expressed as
\begin{equation}\label{trans}
a^{\mu}=a \cdot n\, \bar{n}^{\mu}+a \cdot \bar{n}\, n^{\mu}+a_{\scriptscriptstyle{T}}^{\mu},
\end{equation}
 where $a_{\scriptscriptstyle{T}}^{\mu}$ is the transverse part of the vector.

The matrix element of the \textsf{P}-odd EMT has been parametrized in Ref.~\cite{Lorce:2014mxa} for a spin-1/2 hadron. If we eliminate the polarization-dependent terms, we find that there is only one\footnote{Since $\hat{T}^{\mu\nu}_{q5}(x)$ is a local gauge-invariant operator, its matrix elements must be independent of the light-front vectors $n$ and $\bar n$. This means that one has to set $\tilde C^q(t)=2\tilde F^q(t)$ in the results of~\cite{Tan:2021osk}. We note, in particular, that a factor $1/2$ is missing on the r.h.s. of Eq.~(12) and on the l.h.s. of Eq.~(20) of that paper.} EMT FF in the case of a spin-0 hadron,
\begin{equation}\label{emtsp0}
\begin{aligned}
 \langle p^{\prime} |\hat{T}^{\mu\nu}_{q5}(0)|p\rangle
=i\epsilon^{\mu \nu \Delta P}\tilde F^q(t).
\end{aligned}
\end{equation}
We used the variables $P=\tfrac{1}{2}(p^{\prime}+p)$, $\Delta=p^{\prime}-p$, $t=\Delta^2$, and the notation $\epsilon^{\mu \nu \Delta P} \equiv \epsilon^{\mu \nu \alpha \beta} \Delta_{\alpha}P_{\beta}$ with $\epsilon_{0123}=1$.
The quark kinetic spin-orbit correlation is then given by
\begin{equation}\label{sl-op}
C^q_z=\frac{\epsilon_{{\scriptscriptstyle{T}}\alpha\beta}}{2P^{+}}\left[i\,\frac{\partial}{\partial\Delta_\alpha}\langle p^{\prime}|\hat{T}^{+ \beta}_{q5}(0)|p\rangle\right]_{\Delta=0}=\tilde F^q(0),
\end{equation}
where the transverse Levi-Civita pseudotensor $\epsilon_{\scriptscriptstyle{T}}^{\mu \nu}$ is defined by
$\epsilon_{\scriptscriptstyle{T}}^{\mu \nu}=\epsilon^{\mu \nu \alpha \beta} n_{\alpha}\bar{n}_{\beta}$.

As a result of Lorentz symmetry, the matrix element of any angular momentum operator (orbital or spin) must be proportional to the hadron spin~\cite{Lorce:2015sqe}. Therefore, it is expected that the total quark OAM $L^q_z=L^q_{zR}+L^q_{zL}$ vanishes for a spin-0 hadron. This is confirmed by the explicit parameterization for the hadronic matrix element of the quark (or gluon) EMT~\cite{Kobzarev:1962wt,Pagels:1966zza,Tanaka:2018wea}.
We can then write for the spin-orbit correlation $C^q_z=L^q_{zR}-L^q_{zL}=2L^q_{zR}=-2L^q_{zL}$.

The EMT FF $\tilde F^q(t)$ can in principle be accessed via the twist-3 axial-vector GPD $G^q_2(x, \xi, t)$ in a spin-0 hadron
\begin{equation} \label{sl-pi}
\tilde F^q(t)= -\int dx\, x\, G^q_2(x, \xi, t)
\end{equation}
with  $\xi=-\Delta^+/(2P^+)$. The GPD $G^q_2(x, \xi, t)$ is defined in a $P_{\scriptscriptstyle{T}}^\mu=0$ frame as~\cite{Anikin:2000em,Meissner:2008ay, Belitsky:2005qn}
\begin{equation}\label{GPDt3}
\begin{aligned}
& \frac{1}{2}\int\frac{d z^-}{2\pi}\,e^{ix P^+z^-}\langle p^{\prime}|\overline\psi(-\tfrac{z^-}{2})\gamma_{\scriptscriptstyle{T}}^\mu \gamma_5 \mathcal \psi(\tfrac{z^-}{2})|p\rangle \\
&=\frac{i\epsilon_{\scriptscriptstyle{T}}^{\mu \Delta}}{2P^+}\, G^q_2(x, \xi, t),
\end{aligned}
\end{equation}
where the straight light-front gauge link between the quark fields has been omitted for ease of notation.

There exist a priori also gluonic operators that can be regarded as counterparts of $\hat{T}^{\mu\nu}_{q5}(x)$. For instance, we can consider the following gauge invariant operators
 \begin{equation}
 \label{g5op}
i\tilde{F}^{\mu\alpha}(x)F_\alpha^{\phantom{\alpha}\nu}(x)\quad\text{and}\quad -\tfrac{i}{4}\,g^{\mu\nu}  \tilde{F}^{\alpha\beta}(x)F_{\alpha\beta}(x),
\end{equation}
where  $\tilde{F}^{\mu\nu}= \tfrac{1}{2} \epsilon^{\mu \nu \alpha \beta}F_{\alpha \beta} $ is the dual field. These operators are, in fact, identical owing to the Schouten identity. They have the same parity and time-reversal symmetries as $\hat{T}^{\mu\nu}_{q5}(x)$. The parametrization of their matrix elements should therefore be the same as in Eq.~\eqref{emtsp0}. Since the latter is antisymmetric in $\mu$ and $\nu$ while the operators in Eq.~\eqref{g5op} are symmetric, we conclude that there is no gluonic contribution to the spin-orbit correlation in a spin-$0$ hadron.
Moreover, there is no way to parametrize a symmetric P-odd EMT in the case of a spin-0 target, which means that the gluonic part vanishes. Since the spin-orbit correlation $C_z^q$ is not protected by a symmetry, it depends a priori on the renormalization scale, similarly to the quark spin contribution $\Delta\Sigma$ in a spin-1/2 hadron~\cite{Bass:1992ti}.

Using the QCD equation of motion, one can derive the relation~\cite{Lorce:2014mxa}
\begin{equation}
\label{op-anti}
   \overline\psi \gamma^{[\mu} \gamma_5 i \overset{\leftrightarrow}{D }\!\!\!\!\!\phantom{D }^{\nu ]}  \psi
   = 2m_q \overline\psi i \sigma^{\mu \nu} \gamma_5 \psi - \epsilon^{\mu \nu \alpha \beta} \partial_{\alpha} (\overline{\psi}\gamma_{\beta}\psi ),
\end{equation}
where $a^{[\mu}b^{\nu]}=a^{\mu}b^{\nu}-a^{\nu}b^{\mu}$, and $m_q$ is the quark mass. By taking the matrix element of Eq.~\eqref{op-anti}, one can express the EMT FF $\tilde F^q(t)$ in terms of the vector and tensor FFs for a spin-0 hadron
\begin{equation}\label{eom}
\tilde F^q(t)=\frac{1}{2} \left[-F^q(t)+\frac{m_q}{M}\,H^q(t) \right],
\end{equation}
where $M$ is the hadron mass. Eq.~\eqref{eom} formally coincides with the unpolarized part of the nucleon case~\cite{Lorce:2014mxa}. A similar expression with the FFs replaced by the first Mellin moment of GPDs was found in Ref.~\cite{Tan:2021osk}.
The vector and tensor FFs for the flavor $q$ are defined as~\cite{Hagler:2009ni}

\begin{equation}\label{emffs}
\begin{aligned}
 \langle p^{\prime} | \overline\psi(0)\gamma^\mu \psi(0)|p\rangle&=2P^{\mu}F^q(t), \\
\langle p^{\prime} | \overline\psi(0)i \sigma^{\mu \nu}\gamma_5\psi(0)|p\rangle&=\frac{i\epsilon^{\mu \nu \Delta P}}{M}\,H^q(t).
\end{aligned}
\end{equation}
Interestingly, the matrix element of the divergence of the tensor current vanishes for a spin-$0$ hadron, unlike the spin-$1/2$ case. $F^q(0)$ is the valence quark number for the quark flavor $q$ in a spin-0 hadron. Similarly, one can consider $H^q(0)$ as a new ``charge'' which can be expressed as the first moment of the leading-twist tensor GPD~\cite{Hagler:2009ni}
\begin{equation}\label{sl-pi1}
H^q(t)= 2\int dx\, H^q_1(x, \xi, t),
\end{equation}
where $H^q_1(x, \xi, t)$ is defined by~\cite{Meissner:2008ay, Belitsky:2005qn}
 \begin{equation}\label{GPDt2}
\begin{aligned}
& \frac{1}{2} \int\frac{d z^-}{2\pi}\,e^{ix P^+z^-}\langle p^{\prime}|\overline\psi(-\tfrac{z^-}{2})i \sigma^{i +} \gamma_5 \mathcal \psi(\tfrac{z^-}{2})|p\rangle \\
& =- \frac{i\epsilon_T^{i \Delta}}{M}\, H^q_1(x, \xi, t).
\end{aligned}
\end{equation}
Since the $u$ and $d$ quark masses are much smaller than the hadron mass, we obtain the remarkably simple
relation
\begin{equation}
\tilde F^q(t)=-\tfrac{1}{2}F^q(t)+\mathcal O(\tfrac{m_q}{M}).
\end{equation}

Charge conjugation symmetry implies the following relation between the hadron and anti-hadron vector FFs,
\begin{equation}\label{char}
F^q_{h}(t)=-F^q_{\bar{h}}(t).
\end{equation}
For pions,
the isospin and charge conjugation
symmetries lead us to
\begin{equation}\label{pion}
\begin{aligned}
F_{\pi^+}^u(t)& =-F_{\pi^+}^d(t)= -F_{\pi^-}^u(t)=F_{\pi^{-}}^d(t)=\mathsf{F}_{\pi^+}(t), \\
F_{\pi^0}^q(t)&=F_{\pi^{\pm}}^s(t)=0,
\end{aligned}
\end{equation}
where $\mathsf{F}_{\pi^+}(t)$ is the total electromagnetic (EM) FF for $\pi^+$ that includes the quark electric charges $e_q$
\begin{equation}\label{pion1}
\begin{aligned}
\mathsf{F}_{\pi^+}(t)=\sum_q e_qF^q_{\pi^+}(t).
\end{aligned}
\end{equation}
The notation $\mathsf{F}_S(t)$ applies also to other types of scalar hadrons $S$.
Note, however, that the relations in Eq.~\eqref{pion} can not be used for the kaons due to the different isospin. Instead, one has
\begin{equation}\label{kaon}
F_{K^+}^{u+d}(t) =F_{K^0}^{u+d}(t),\qquad
F_{K^+}^{u-d}(t) =-F_{K^0}^{u-d}(t),
\end{equation}
and similar relations can be derived for $K^-$ and $\bar{K}^0$ using Eq.~\eqref{char}.
In addition to scalar mesons, one can also investigate the EM FFs of the scalar nuclei such as $\alpha$ particle and anti-$\alpha$ particle using the isospin symmetry,
\begin{equation}\label{al-al}
\begin{aligned}
F_{\alpha}^{u}(t) &=F_{\alpha}^{d}(t)=\tfrac{1}{3}\mathsf{F}_{\alpha}(t),\\
F_{\bar{\alpha}}^{u}(t) &=F_{\bar{\alpha}}^{d}(t)=\tfrac{1}{3}\mathsf{F}_{\bar{\alpha}}(t),
\end{aligned}
\end{equation}
where the strange quark contribution is neglected.
Since the $\alpha$ particle is composed of nucleons,
its EM FF~\cite{Schiavilla:1990zz} can also be calculated from the nucleon EM FFs using the Argonne two-nucleon~\cite{Wiringa:1984tg} and Urbana-VII three-nucleon interactions~\cite{Schiavilla:1985gb,Arnold:1978qs}.

In the forward limit, we conclude that the spin-orbit correlation is approximately equal to minus half of the valence quark number for the flavor $q$ in a spin-0 hadron,
\begin{equation}\label{cz}
C^q_z=-\tfrac{1}{2}F^q(0),
\end{equation}
where the quark mass-dependent term is neglected. We therefore expect in general a very mild renormalization scale dependence for $C^q_z$. The spin-orbit correlation for a single quark flavor does not exist in scalar particles with fixed charge parity such as $\pi^0$ and $f_0$ mesons. In the case of a scalar meson, the spin-orbit correlation vanishes for the sum over quark flavors
\begin{equation}
C_z \equiv \sum_q C^q_z=0.
\end{equation}
This convention for the sum over flavors applies to the entire work. However, one finds $C_z=-6$ and $C_z=6$ for the $\alpha$ and anti-$\alpha$ particles, respectively.

\section{2D and 3D spatial distributions}
\label{sect3}

In the Breit frame, 3D spatial distributions of quarks and gluons that reveal the mechanical properties in hadrons can be obtained from the static EMT~\cite{Polyakov:2002yz,Polyakov:2018zvc,Lorce:2018egm,Burkert:2023wzr}. Similarly, the \textsf{P}-odd EMT allows us to define the 3D spatial distributions for left- and right-handed quarks, separately. In the Breit frame, there is no energy transfer between the incoming and outgoing hadrons so that
\begin{equation}\label{bfkin}
\begin{aligned}
p^{\mu}=(E, -\tfrac{\vec{\Delta}}{2}),\qquad
p^{\prime \mu}=(E,\tfrac{\vec{\Delta}}{2}),
\end{aligned}
\end{equation}
with $E=\sqrt{M^2+\vec{\Delta}^2/4}$.
The spatial distributions are then defined by a Fourier transform with respect to $\vec{\Delta}$
\begin{equation}\label{for0}
\begin{aligned}
\mathcal T_{q5}^{ij}(\vec{r}) &=\int \frac{d^3\Delta}{(2\pi)^3}\,  e^{-i \vec{\Delta}\cdot \vec{r}  } \,   \frac{\langle p^{\prime} |\hat{T}^{ij}_{q5}(0)|p\rangle}{2E}  \\
&=\epsilon^{ijk} e_r^k\, v_q(r),
\end{aligned}
\end{equation}
where $e_r^i= r^i/|\vec r|$, $\epsilon^{ijk}$ is the three-dimensional Levi-Civita symbol, and $v_q(r)$ is a new spatial distribution  associated with  $\tilde F^q(t)$.
The tensor $\mathcal T_{q5}^{ij}(\vec{r})$ is conserved since $\nabla^i (\epsilon^{ijk} e_r^k v_q(r))=0$.
Defining the Fourier transform of $\tilde F^q(t)$ as
\begin{equation}
w_q(r)=\int \frac{d^3\Delta}{(2\pi)^3 }  \,e^{-i \vec{\Delta}\cdot \vec{r}  }\, \tilde F^q(t)
\end{equation}
with $t=-\vec \Delta^2$, we find that
\begin{equation}
e_r^k\, v_q(r)=\tfrac{1}{2}\,\nabla^k_r w_q(r).
\end{equation}
It follows in particular that
\begin{equation}\label{for1}
\int d^3r\,r\,v_q(r)=-\tfrac{3}{2}\tilde F^q(0),
\end{equation}
provided that surface terms vanish.

\begin{figure}[ht]
       \includegraphics[scale=0.3]{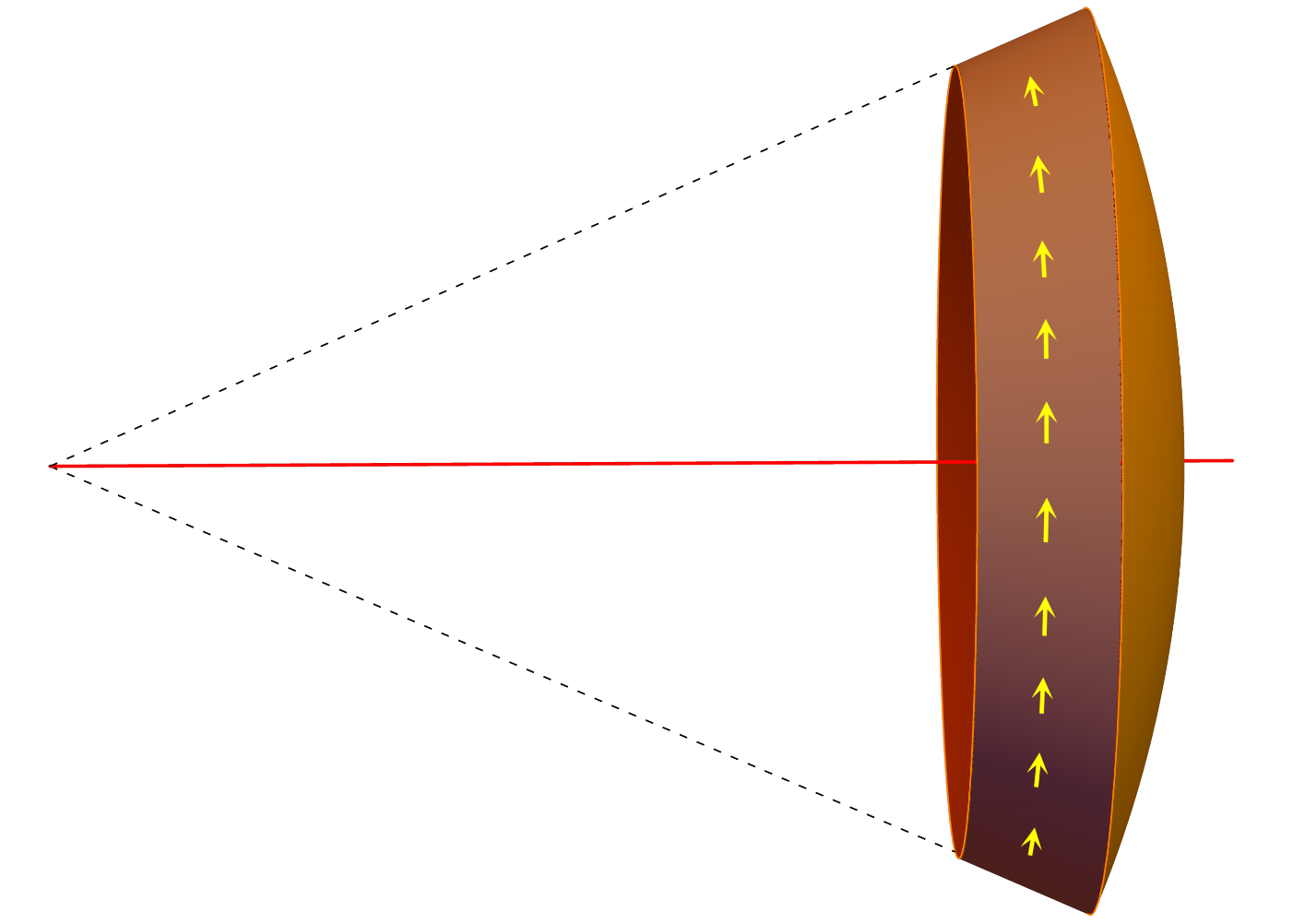}
\caption{Illustration of the chiral stress distribution $v_q(r)$ (here assumed positive) on a spherical cap at a distance $r$ from the center of the hadron.}
\label{fig01}
\end{figure}

Thus, the 3D spatial distributions for left- and right-handed quarks are given by
\begin{equation}\label{for5}
\begin{aligned}
&\mathcal T_{R/L}^{ij} (\vec r)=\frac{1}{2}\left[\mathcal T_q^{ij}(\vec{r}) \pm  \mathcal T_{q5}^{ij}(\vec{r}) \right] \\
&=\frac{1}{2}\left[ (e_r^ie_r^j-\tfrac{1}{3}\delta^{ij})\,s_q(r)+ \delta^{ij} p_q(r) \pm \epsilon^{ijk} e_r^k v_q(r) \right] .
\end{aligned}
\end{equation}
$p_q(r)$ is the isotropic pressure and $s_q(r)$ is the pressure anisotropy (or shear forces)~\cite{Polyakov:2002yz,Polyakov:2018zvc}. The new function $v_q(r)$ does not contribute to the radial force because of parity symmetry. It can be interpreted as a chiral stress, in the sense of a torque field about the radial axis that results from the spin-orbit correlations, see Fig.~\ref{fig01}. Indeed, the component $i$ of the torque acting on a surface with unit normal vector $\vec e_l$ is given by
\begin{equation}
    \mathcal C^{il}_{R,L}(\vec r)=\epsilon^{ijk}r^j\mathcal T^{lk}_{R,L}(\vec r)=\mp \,\frac{1}{2}\left(\delta^{il}-e_r^ie_r^l\right)r\, v_q(r).
\end{equation}
Note that this torque has no effect on surfaces orthogonal to the radial vector. We then define the radial torque as
\begin{equation}
    \tau_{q,r}(r)\equiv\left(\delta^{il}-e_r^ie_r^l\right)\mathcal C^{il}_{R}(\vec r)=-r\, v_q(r).
\end{equation}

It is challenging to access $\tilde F^q(t)$ experimentally
 since it is a higher-twist effect, as indicated by Eq.~\eqref{sl-pi}. However, if we neglect the quark mass, then $\tilde F^q(t)$ is just proportional to the EM FF, for which there are numerous experimental and theoretical studies, as discussed in Sect.~\ref{sect1}.
One can then approximately write $v_q(r)=-\tfrac{1}{4}d\rho_q(r)/dr$ with the charge density $\rho_q(r)$ given by
\begin{equation}\label{charge}
\rho_q(r)=\int \frac{d^3\Delta}{(2\pi)^3 } \, e^{-i \vec{\Delta}\cdot \vec{r}  }\,  F^q(t).
\end{equation}

In Ref.~\cite{Cui:2021aee}, the monopole, dipole and Gaussian functional forms are adopted to describe the pion EM FF of Eq.~\eqref{pion},
\begin{equation}
\label{PiFF}
\mathsf{F}_{\pi^+}(t) =
\left\{
\begin{array}{lr}
\frac{1}{1 - r_\pi^2 t/6} & \mbox{Monopole} \\[1ex]
\frac{1}{(1 - r_\pi^2 t/12)^2} & \mbox{Dipole} \\[1ex]
{\rm e}^{r_\pi^2 t/6} & \mbox{Gaussian}
\end{array}\right.
\end{equation}
where $r_\pi=0.659\pm0.004$ fm is the pion charge radius~\cite{ParticleDataGroup:2020ssz}. The monopole form is consistent with the prediction of Ref.~\cite{Lepage:1979zb}, where perturbative QCD suggests that $\mathsf{F}_{\pi^+}(t) \sim (-t)^{-1}$ at large momentum transfer. However, the simple monopole form leads to a singular charge density $\rho(r)$  at $r=0$~\cite{Miller:2009qu}, and so a similar situation will apply to the chiral stress distribution $v(r)$.
On the other hand, the Drell-Yan-West relation suggests that $\mathsf{F}_{\pi^+}(t) \sim (-t)^{-3/2}$ at large $t$~\cite{Alberg:2024svo}.
It should be noted that the experimental studies of the scalar meson EM FFs focused so far on the low-$t$ region. However, the situation is about to improve. For example, the ongoing measurements of the EM FFs of  pions and kaons can reach the region of $Q^2=-t \sim 6$ GeV$^2$ at the JLAB 12 GeV Continuous Electron Beam Accelerator Facility (CEBAF)~\cite{Dudek:2012vr, Arrington:2021alx}.
In the near future, it will also be possible to measure the meson EM FFs in a much higher $Q^2$-region at the Electron-Ion Collider (EIC) in the US~\cite{Aguilar:2019teb, Arrington:2021biu} and the Electron-ion collider in China (EicC)~\cite{Anderle:2021wcy}. We therefore follow Ref.~\cite{Cui:2021aee} and consider here the three different functional forms for the pion EM FF.
\begin{figure}[htb]
       \includegraphics[scale=0.5]{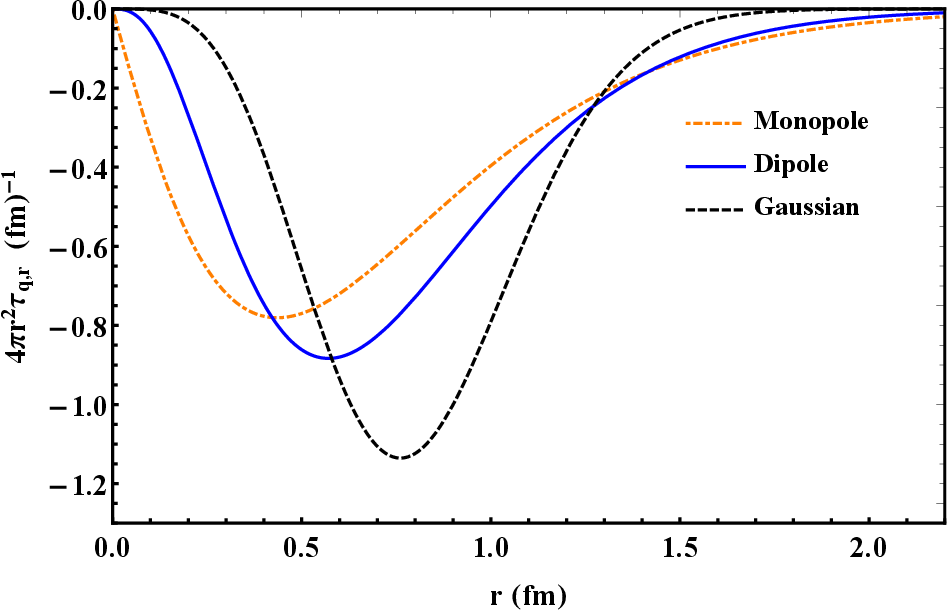}
	\caption{Illustration of torque distributions for the quark flavor $u$ (or the flavor $d$ by isospin symmetry relation) in $\pi^+$  using three different functional forms for the pion EM FF.}
	\label{fig02}
\end{figure}

In Fig.~\ref{fig02}, we use the pion EM FF of Eq.~\eqref{PiFF}
to illustrate the torque distributions $4\pi r^2\tau_{q,r}(r)$
for quark flavor $u$ in $\pi^+$ (see Eq.~\eqref{pion} for the relations between different quark flavors). As discussed above, the monopole form will introduce a  singularity of $\tau_{q,r}(r)\sim 1/r$ at the center,  in contrast to the dipole and Gaussian forms of the pion EM FF.  The torque distributions   are concentrated in  the region of $0.3-1.0$ fm for all three forms.

Following the spirit of Ref.~\cite{Lorce:2018egm}, we can define the two-dimensional chiral stress distribution in the elastic frame, where $\vec{\Delta}$ lies in the plane transverse to $\vec{P}=P_z\vec{e}_z$, via
\begin{equation}\label{for2d}
\begin{aligned}
\mathcal T_{q5}^{ij}(\vec{b}_{\scriptscriptstyle{T}}) &=\int \frac{d^2\Delta_{\scriptscriptstyle{T}}}{(2\pi)^3}\,  e^{-i \vec{\Delta}_{\scriptscriptstyle{T}}\cdot \vec{b}_{\scriptscriptstyle{T}}  } \,   \frac{\langle p^{\prime} |\hat{T}^{ij}_{q5}(0)|p\rangle}{2P^0}\\
&=\epsilon^{ijk} \int d r_z\,e_r^k\, v_q(r)
\end{aligned}
\end{equation}
with $\vec r=(\vec b_{\scriptscriptstyle{T}},r_z)$. Clearly, $\mathcal T_{q5}^{ij}(\vec{b}_{\scriptscriptstyle{T}})$ vanishes when $i,j$ are both transverse indices. Interestingly, the non-trivial components $\mathcal T_{q5}^{3i}(\vec{b}_{\scriptscriptstyle{T}})=-\mathcal T_{q5}^{i3}(\vec{b}_{\scriptscriptstyle{T}})$ do not depend on $P_z$ and are therefore the same in both the transverse Breit frame ($P_z\to 0$) and the infinite-momentum frame ($P_z\to\infty$). Similarly, in the symmetric Drell-Yan (DY) frame characterized by $\Delta^+=0$ and $\vec{P}_T=0$, the non-trivial chiral stress distribution on the light front is obtained from
\begin{equation}
    \mathcal T_{q5}^{+i}(\vec{b}_{\scriptscriptstyle{T}}) =\int \frac{d^2\Delta_{\scriptscriptstyle{T}}}{(2\pi)^3}\,  e^{-i \vec{\Delta}_{\scriptscriptstyle{T}}\cdot \vec{b}_{\scriptscriptstyle{T}}  } \,   \frac{\langle p^{\prime} |\hat{T}^{+i}_{q5}(0)|p\rangle}{2P^+}.
\end{equation}
Like in Refs.~\cite{Lorce:2017wkb,Lorce:2018egm}, we can also define the distribution of the kinetic spin-orbit correlation in impact-parameter space as follows:
 \begin{equation}\label{sl-dis}
 \begin{aligned}
 \langle \hat{C}^q_z\rangle(b_{\perp})&=\epsilon^{jk}_T{b}^j_{\scriptscriptstyle{T}}\mathcal T_{q5}^{+k}(\vec{b}_{\scriptscriptstyle{T}})\\
 &= \frac{-i\epsilon^{jk}_T}{2P^+} \int \frac{d^2\Delta_{\scriptscriptstyle{T}}}{(2\pi)^2 } \, e^{-i \vec{\Delta}_{\scriptscriptstyle{T}}\cdot \vec{b}_{\scriptscriptstyle{T}}  }  \,\frac{\partial  T^{+k}_{q5} }{\partial \Delta^j_T} \Bigg|_{\text{DY}} \\
 &=\int \frac{d^2\Delta_{\scriptscriptstyle{T}}}{(2\pi)^2 }\,  e^{-i \vec{\Delta}_{\scriptscriptstyle{T}}\cdot \vec{b}_{\scriptscriptstyle{T}}  } \left[ \tilde F^q(t)+t\frac{d\tilde F^q(t)}{dt} \right]
 \end{aligned}
\end{equation}
with $b_\perp=|\vec{b}_{\scriptscriptstyle{T}}|$.

\begin{figure}[ht]
       \includegraphics[scale=0.5]{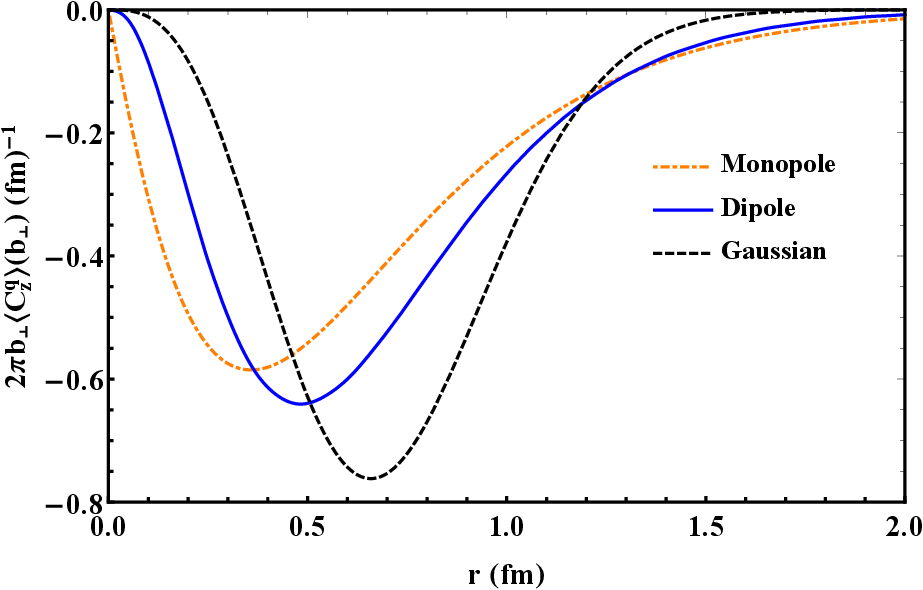}
	\caption{Distributions of the kinetic spin-orbit correlation for the quark flavor $u$ (or $d$) in the impact-parameter space using three different functional forms for the pion EM FF. }
	\label{fig03}
\end{figure}

In Fig.~\ref{fig03}, we illustrate the distributions of the kinetic spin-orbit correlation in impact-parameter space for the quark flavor $u$ (or $d$) in $\pi^+$,  using the monopole, dipole and Gaussian functional forms for the pion EM FF in Eq.~\eqref{PiFF}. The area enclosed by the each curve is the spin-orbit correlation of $\pi^+$,
$C^u_z=-1/2$, as indicated by Eq.~\eqref{cz}. The three curves are quite different at small $b_{\perp}$, but they all indicate that the distribution of the spin-orbit correlation concentrates in the region of $b_{\perp}<1$ fm.

\section{Summary}
\label{sect5}
It is well known that spin-0 hadrons do not admit non-trivial (orbital or spin) angular momentum  contributions. However, their spin structure can still be characterized by the spin-orbit correlation, which measures the difference of orbital angular momentum between right- and left-handed partons. We showed that the quark spin-orbit correlation is described by the \textsf{P}-odd energy-momentum tensor form factor (FF) $\tilde F^q(t)$, which can be expressed as the second moment of the twist-3 axial-vector generalized parton distribution, in principle accessible through deeply-virtual Compton scattering (DVCS) experiments. However, such measurements are challenging due to the suppression of the cross section by $\alpha^2$ and higher-twist effects, without mentioning the absence of meson targets.
Fortunately, $\tilde F^q(t)$ can alternatively be expressed in terms of the vector (or electromagnetic) and tensor FFs using the QCD equation of motion. If one neglects the quark mass contribution, the quark spin-orbit correlation is then equal to minus half of the valence quark number in spin-0 hadrons. We studied also the spatial distribution of the \textsf{P}-odd stress tensor and found that the spin-orbit correlation leads to a new contribution to the pressure inside spin-0 (and hence also higher-spin) hadrons, which we interpreted as chiral stress. Finally, we introduced the distribution of spin-orbit correlation in impact-parameter space.

 In contrast to the delicate DVCS measurements, the vector FF $F^q(t)$ of spin-0 hadrons can more easily be measured in electron-hadron elastic scattering, and has already been extracted to some extent in the case of pions, kaons, and $\alpha$ particles. We used the available pion results to illustrate our results. In the near future, extensive measurements of the vector FFs of spin-0 hadrons will be conducted  at various facilities such as JLab 12 GeV CEBAF and the Electron-Ion Colliders in the US and China. Our study shows that these FFs provide key physical insights into the internal structure of scalar hadrons, which go well beyond the sole electromagnetic structure.

\section*{Acknowledgements}
Qin-Tao Song expresses gratitude for the hospitality during his stay at CPHT in \'Ecole polytechnique.
Qin-Tao Song was supported by the National Natural Science Foundation of
China under Grant Number 12005191.

\end{document}